\documentclass[%
 reprint,
 amsmath,amssymb,
 aps,
]{revtex4-2}
\usepackage{float} 
\usepackage{graphicx}
\usepackage{dcolumn}
\usepackage{bm}
\usepackage{amsmath}
\usepackage{bbm}
\usepackage{bbold}
\usepackage{braket}
\usepackage{array}
\usepackage{xcolor}

\begin{document}

\title{Spectroscopic signatures of tetralayer graphene polytypes}

\author{Andrew McEllistrim}
 \email{andrew.mcellistrim@postgrad.manchester.ac.uk}
 \affiliation{National Graphene Institute, University of Manchester, Booth Street East, Manchester M13 9PL, UK}
\affiliation{Department of Physics and Astronomy, University of Manchester, Oxford Road, Manchester, M13 9PL, UK}

\author{Aitor Garcia-Ruiz}
 \affiliation{National Graphene Institute, University of Manchester, Booth Street East, Manchester M13 9PL, UK}
\affiliation{Department of Physics and Astronomy, University of Manchester, Oxford Road, Manchester, M13 9PL, UK}

\author{Zachary A. H. Goodwin }
 \affiliation{National Graphene Institute, University of Manchester, Booth Street East, Manchester M13 9PL, UK}
\affiliation{Department of Physics and Astronomy, University of Manchester, Oxford Road, Manchester, M13 9PL, UK}
\affiliation{John A. Paulson School of Engineering and Applied Sciences, Havard University, Cambridge, MA, United States, 02138}

\author{Vladimir I. Fal'ko}
 \affiliation{National Graphene Institute, University of Manchester, Booth Street East, Manchester M13 9PL, UK}
\affiliation{Department of Physics and Astronomy, University of Manchester, Oxford Road, Manchester, M13 9PL, UK}
 \affiliation{Henry Royce Institute for Advanced Materials, University of Manchester, Oxford Road, Manchester, M13 9PL, UK}
\date{14/02/2023}

\begin{abstract}
Tetralayer graphene has recently become a new addition to the family of few-layer graphenes with versatile electronic properties. This material can be realised in three distinctive stacking configurations, for which we determine spectroscopic signatures in angle-resolved photoemission spectroscopy (ARPES), dynamical optical conductivity, and Raman spectra of inter-band excitations. The reported library of spectral features of tetralayer graphenes can be used for the non-invasive identification of the stacking order realised in a particular film.  
\end{abstract}

\maketitle


\section{Introduction}
While graphene has already been the subject of numerous experimental and theoretical studies, the family of few-layer graphene polytypes is still growing, showing various stacking-dependent electronic properties. In this respect, studies of bilayers and trilayers have received most of the attention, but, recently, the family of graphene polytypes got a new member: tetralayer graphene \cite{Mak2010,Konstantin_Experimental_2022,Grushina2015, Yoon2017, Myhro2018, Shi2018, Abderrahim_Band_2019, Che2020, Liang2022, Chou2022, Pierucci2016}. For tetralayer graphene (4LG), there are four possible interlayer stacking configurations with Bernal (AB or BA) stacking of adjacent layers. Two of those are related by inversion, leaving only three unique structures with their specific electronic band structures: ABAB (Bernal 4LG), ABCA (rhombohedral 4LG), and an allusive ABCB or ABAC (mixed-stacking 4LG) \cite{wirth_experimental_2022}. The latter, mixed-stacking 4LG is an interesting system, as it appears to be the thinnest few-layer graphene that has neither inversion nor mirror plane (z$\rightarrow$-z) symmetry, which allows it to possess a spontaneous out-of-plane electric polarisation. To mention, all these four structures can be obtained from each other by shear displacements of the layer \cite{Mishchenko2020}, and they also appear in pairs across moiré patterns in flakes of small-angle-twisted tetralayers, produced by mechanically transferring a monolayer on a trilayer \cite{Goodwin2021}, or by assembling together to aligned bilayer flakes \cite{shi_electronic_2020,Tomic_Scattering_2020}. All of this leads to the necessity to develop a set of non-invasive characterisation methods to identify local stacking orders in tetralayer graphenes.

In this paper, we analyse spectroscopic responses of tetralayer graphene systems from angle-resolved photo-electric spectroscopy (ARPES), optical visibility in the infrared (IR) range and electronic Raman scattering. The below-reported results are obtained using a hybrid $\boldsymbol{k\cdot p}$ theory-tight-binding approach, described in section II, based on a fully-parametrised Slonczewski-Weiss-McClure (SWMcC) model of graphite\cite{Slonczeswki_Band_1958,McClure_Band_1957,McClure_Theory_1960,Aitor_Full_2021}, with each of the spectroscopic characteristics discussed separately in sections III, IV, and V, respectively.
\vspace{3mm}
\section{Band structure of tetralayer graphenes}\label{Sec:band_structure}
Tetralayer graphene can exist in four types of stacking orders, depending on the relative shifts between adjacent layers. In the Bernal configuration, the carbons atoms in layers 1 and 3 occupy identical positions, while they are laterally shifted by $\boldsymbol{\tau}=(0,a/\sqrt{3})$ with respect to layers 2 and 4, $a \approx 2.46$~{\AA} being the lattice constant of graphene. In the rhombohedral configuration, each layer is shifted by $\boldsymbol{\tau}$ with respect to the top adjacent layer. Tetralayer graphite allows for another type of configuration that features sequences of Bernal and rhombohedral arrangement. In this mixed configuration, also known as ABCB (ABAC), the graphene layer 4 (3) lie directly on top of layer 2 (1), while the first and third (second and fourth) are shifted by $\boldsymbol{\tau}$ and $-\boldsymbol{\tau}$, respectively. 
\begin{figure*}
    \centering
    \includegraphics[width=1\textwidth]{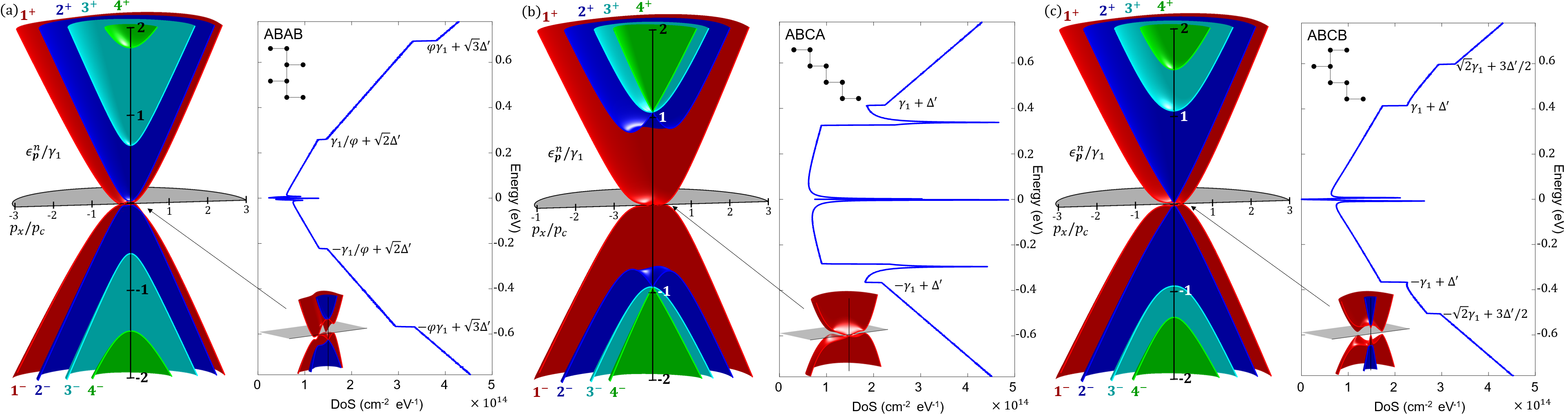}
    \caption{Band structure and density of states (DoS) of (a) Bernal, (b) rhombohedral and (c) mixed tetralayer graphene. The insets at the bottom left of the DoS show the low energy band structure within an energy window of $\pm30$ meV. The effect of $\Delta'$ becomes apparent at the edges of the bands $3^{\pm}$ and $4^{\pm}$ at $\boldsymbol{p}=\boldsymbol{0}$, which induce steps in the DoS. In the labels of panel (a) ${p_c}=\gamma_1/v$ and $\varphi\equiv\frac{1+\sqrt{5}}{2}$.}
    \label{Fig:band_structures}
\end{figure*}

Electronic properties for each of these configurations can be determined using the hybrid $\boldsymbol{k} \cdot \boldsymbol{p}$ theory - tight binding Hamiltonians \cite{Slonczeswki_Band_1958,McClure_Band_1957,McClure_Theory_1960}:
\begin{widetext}
\begin{subequations}\label{Eq:Ham}
\begin{align}
&\mathcal{H}_{\mathrm{ABAB}}=
\left(
\begin{matrix}
H_g+\frac{\Delta'}{2}(\mathbb{1}_2+\sigma_z)&
V_{\mathrm{AB}}&
W_{\mathrm{ABA}}&
0\\
V_{\mathrm{AB}}^\dagger&
H_g+\Delta'(\mathbb{1}_2-\sigma_z)&
V_{\mathrm{AB}}^\dagger&
W_{\mathrm{BAB}}\\
W_{\mathrm{ABA}}^\dagger&
V_{\mathrm{AB}}&
H_g+\Delta'(\mathbb{1}_2+\sigma_z)&
V\\
0&
W_{\mathrm{BAB}}^\dagger&
V_{\mathrm{AB}}^\dagger&
H_g+\frac{\Delta'}{2}(\mathbb{1}_2-\sigma_z)
\end{matrix}
\right),\\
&\mathcal{H}_{\mathrm{ABCA}}=
\left(
\begin{matrix}
H_g+\frac{\Delta'}{2}(\mathbb{1}_2+\sigma_z)&
V_{\mathrm{AB}}&
W_{\mathrm{ABC}}&
0\\
V_{\mathrm{AB}}^\dagger&
H_g+\Delta'\mathbb{1}_2
&V_{\mathrm{AB}}&
W_{\mathrm{ABC}}\\
W_{\mathrm{ABC}}^\dagger&
V_{\mathrm{AB}}^\dagger&
H_g+\Delta'\mathbb{1}_2&
V\\
0&
W_{\mathrm{ABC}}^\dagger&
V_{\mathrm{AB}}^\dagger&
H_g+\frac{\Delta'}{2}(\mathbb{1}_2-\sigma_z)
\end{matrix}
\right),\\
&\mathcal{H}_{\mathrm{ABCB}}=
\left(
\begin{matrix}
H_g+\frac{\Delta'}{2}(\mathbb{1}_2+\sigma_z)&
V_{\mathrm{AB}}&W_{\mathrm{ABC}}&0\\
V_{\mathrm{AB}}^\dagger&
H_g+\Delta'\mathbb{1}_2&
V_{\mathrm{AB}}&
W_{\mathrm{ABA}}\\
W_{\mathrm{ABC}}^\dagger&
V_{\mathrm{AB}}^\dagger&
H_g+\Delta'(\mathbb{1}_2+\sigma_z)&
V_{\mathrm{AB}}^\dagger\\
0&W_{\mathrm{ABA}}^\dagger&
V_{\mathrm{AB}}&
H_g+\frac{\Delta'}{2}(\mathbb{1}_2+\sigma_z)
\end{matrix}
\right),\\
&\mathcal{H}_{\mathrm{ABAC}}=
\left(
\begin{matrix}
H_g+\frac{\Delta'}{2}(\mathbb{1}_2+\sigma_z)&
V_{\mathrm{AB}}&
W_{\mathrm{ABA}}&0\\
V_{\mathrm{AB}}^\dagger&
H_g+\Delta'(\mathbb{1}_2-\sigma_z)&
V_{\mathrm{AB}}&W_{\mathrm{ABC}}^\dagger\\
W_{\mathrm{ABA}}^\dagger&
V_{\mathrm{AB}}^\dagger&
H_g+\Delta'\mathbb{1}_2&
V_{\mathrm{AB}}^\dagger\\
0&W_{\mathrm{ABC}}&
V_{\mathrm{AB}}&
H_g+\frac{\Delta'}{2}(\mathbb{1}_2+\sigma_z)
\end{matrix}
\right),\\
&H_g=
\left(
\begin{matrix}
0& v \hbar \pi_{\xi}^*\\
v \hbar \pi_{\xi}&0
\end{matrix}
\right),\,
V_{\mathrm{AB}}=\left(
\begin{matrix}
-v_4 \pi_{\xi}& \gamma_1\\
-v_3 \pi_{\xi}^*& -v_4 \pi_{\xi}
\end{matrix}
\right),\,
\pi_{\xi}\approx \xi p_x + i p_y\\
&W_{\mathrm{ABA}}=
\left(
\begin{matrix}
\frac{\gamma_5}{2}&0\\
0&\frac{\gamma_2}{2}
\end{matrix}
\right),\,
W_{\mathrm{BAB}}=
\left(
\begin{matrix}
\frac{\gamma_2}{2}&0\\
0&\frac{\gamma_5}{2}
\end{matrix}
\right),\,
W_{\mathrm{ABC}}=
\left(
\begin{matrix}
0&0\\
\frac{\gamma_2}{2}&0
\end{matrix}
\right),\,
\sigma_z=
\left(
\begin{matrix}
    1&0\\0&-1
\end{matrix}
\right).
\nonumber
\end{align}
\end{subequations}
\end{widetext}
Here $\mathbb{1}_2$ is the $2\times2$ matrix and $\boldsymbol{p}=(p_x,p_y)$ is the momentum measured from the corners of the Brillouin zone, $K_{\xi}=\frac{4\pi}{3a}(\xi,0)$. In the above equations, we use the following values \cite{Yin2019, sergey_param} for the coupling parameters: $(v,v_3,v_4)=(1,0.1,0.022)10^6$ m/s and $(\gamma_1,\gamma_2,\gamma_5,\Delta')=(390,-17,38,25)$ meV.

The band structure of all of these systems can be calculated by diagonalising the above Hamiltonians. Within the 3 polytypes of 4LG, each band structure has defining characteristics that can be related to the density of states (DoS). Bernal 4LG is unique as it has the lowest energy appearance of its $3^\pm$ band edges at $\textit{\textbf{p}}$=\textbf{0}, seen as a step at $\gamma_1/\varphi + \sqrt{2}\Delta'$ meV in the DoS, where $\varphi$ is the golden number.  It also has a comparatively large gap between this and its $4^\pm$ band edge. Rhombohedral is the only polytype without a Bernal stack in its configuration, thus the only structure with just bands $1^\pm $ in its low energy dispersion (see large van Hove singularity (vHs) at 0 meV). Additionally, rhombohedral contains a Mexican hat-like dispersion at an energy of $\pm \gamma_1+\Delta'$, where the minimum of the band edge is off the $\textit{\textbf{p}}$=\textbf{0} line. The mixed stacking order has features of both Bernal and rhombohedral graphenes. Firstly it has 2 pairs of degenerate bands in its low energy dispersion ($ 1^\pm,2^\pm$) like Bernal and it also has its $3^\pm $ band edge emerging at $\pm\gamma_1+\Delta'$ like in rhombohedral graphene \cite{Kuzmenko_determination_2009}. This commonality in the $ 1^\pm,2^\pm$ bands with Bernal is mirrored by a small vHs at 0 meV in the DoS. The position of the $3^\pm $ band can be seen as a step in the DoS plots coinciding with the vHs of rhombohedral at the same energy.

\section{ARPES spectra of 4LGs}

In ARPES experiments, a sample is irradiated by a source of high-energy photons ($ \omega \approx 60-100$ eV), which eject electrons due to the photoelectric effect \cite{Einstein1905}. Because it is possible to trace back the energy and momentum of the scattered electrons, this technique allows us to reconstruct the dispersion of materials, and due to its surface sensitivity, it has been widely employed to study two-dimensional physics, such as transition-metal-dichalcogenides \cite{Sayers_Coherent_2020}, surface states in topological insulators \cite{Ryu_Topological_2021,Baringthon_Topological_2022} and graphene systems \cite{Thompson_2020}. 

Following previous works \cite{MuchaKruczyski2008}, we compute ARPES as,
\begin{widetext}
    \begin{align}\label{Eq:ARPES}
\mathcal{A}^{n}
(E,{\boldsymbol{p}})=
\left|
\sum_{\lambda=0}^3
\sum_{s}
F^\lambda
\int d\boldsymbol{r}
e^{-\frac{i}{\hbar}
\left(
\hbar\boldsymbol{K}_\xi^{(0)}+\boldsymbol{p}_\mathrm{w}
\right)\cdot\boldsymbol{r}}
\psi_{\lambda_s}^{n}(\boldsymbol{p})
\right|^2 
\frac{f(E-\epsilon_{\boldsymbol{p}}^{n})\Gamma/\pi}{(\epsilon_{\boldsymbol{p}}^{n}-E)^2+\Gamma^2}.
\end{align}
\end{widetext}

Here, $\psi_{\lambda_{s}}^{n}(\boldsymbol{p})$ is multi-component wave function composed of the electron wave amplitudes on $s=$ A/B sublattices in each layer, $\lambda$, of a  Bloch state at the point $\hbar\boldsymbol{K}_\xi^{(0)}+\boldsymbol{p}_\mathrm{w}$, in the Brillouin zone of graphene. Indices $n^\mathrm{th}$ identify valence ($n^-$) and conduction ($n^+$) bands in the few-layer grapheen spectrum, and  $f(E-\epsilon_{\boldsymbol{p}}^{n})$ is Fermi distribution. We also introduce a factor $F=|F| e^{\frac{i}{\hbar}p_z\cdot c_0}$, where $|F|=0.4$ accounts for the attenuating effect of ejecting electrons from layers further away from the top layer ($\lambda=0$), and a
the phase factor accounts for the mutual phase shifts acquired by the partial waves of electrons originating(with an out-of-plane momentum, $p_z$) in different layers (we use $c_0=3.35 \mathrm{\AA}$ as the interlayer distance in graphitic films). Equation \ref{Eq:ARPES} states that the observable ARPES intensity is determined by the interference of electron waves originating from  the A and B sublattices of each of the four layers, and it also takes into account inelastic broadening and  spectrometer resolution, $\Gamma=60$ meV.

\vspace{3mm}
In Fig. 2a, we show three sets of ARPES maps for Bernal, rhombohedral and mixed-stacking 4LG's. The set of four images on the top show momentum cuts through the Kappa point in the $k_y$ direction. These images directly coincide with the band structures presented in Fig.\ref{Fig:band_structures}. All other images are the constant energy maps on the $p_x-p_y$ (around $K$-point) which reflect the layer and sublattice composition of the electronic wave functions. Their appearance in ARPES depends on the energy, $\omega$, of the incoming photon, as that determines the out-of-plane momentum of the excited electrons, $p_z=\sqrt{2m(\omega-A)+\epsilon_c-(\hbar K)^2}$, where $\epsilon_c$ is the energy the ARPES measurement was taken at. We estimate that 
$\omega\approx 70$eV corresponds to a $4\pi$ phase shift for electron waves originating from two consecutive monolayers, and $\omega\approx 100$eV to a $5\pi$ phase shift. The corresponding ARPES patterns for these two choices of photon energies are shown in rows 2-5 of Fig. 2 for constant energy maps at 250 meV and 500 meV below the Fermi level in the material. 

\begin{figure*}
    \centering
    \includegraphics[width=2\columnwidth]{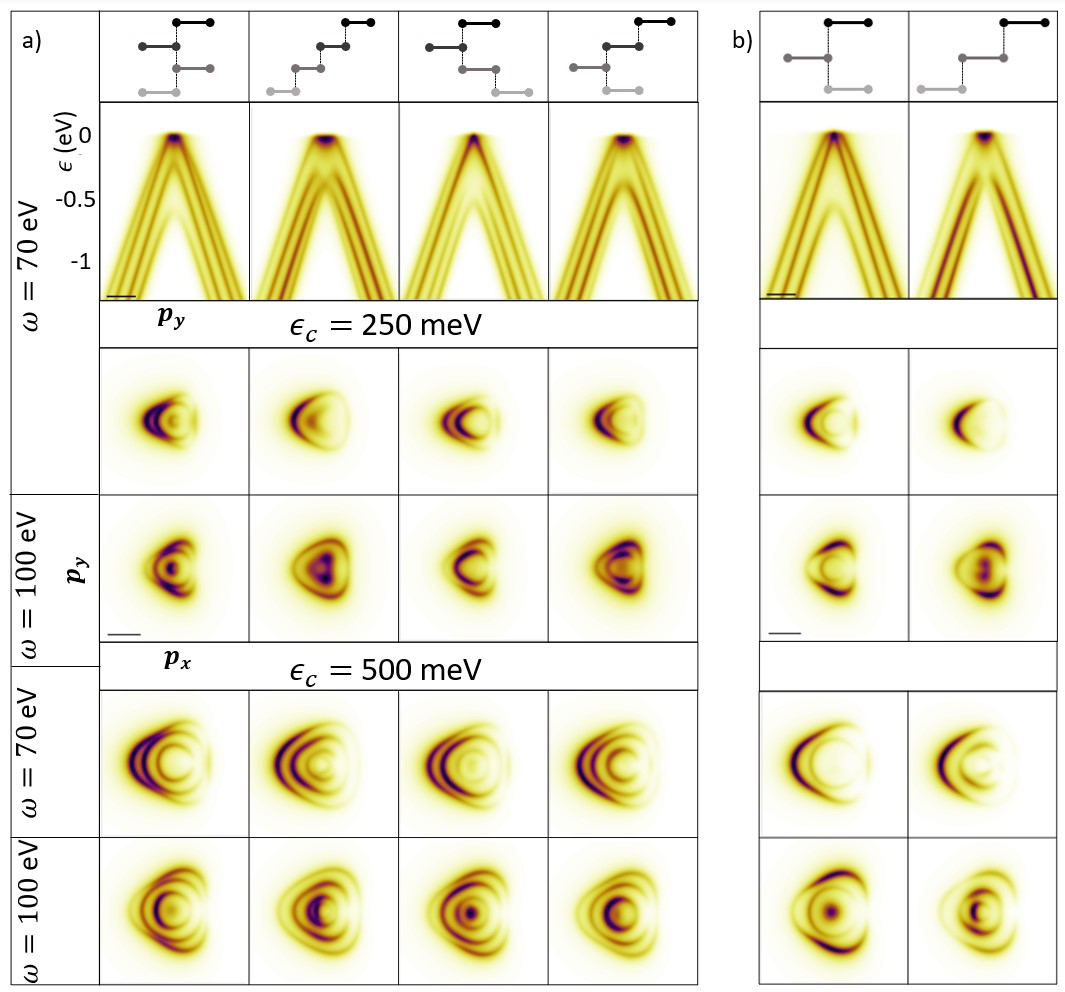}
    \caption{\textbf{a) 4LG spectra, b) 3LG sprectra.} Row 1: Constant Momentum cuts through the $K_{+}$ point, $\omega$=70 eV. Row 2(3), constant energy maps at $\epsilon_c$=-250 meV and $\omega$=70(100) eV. Row 4(5), constant energy maps at $\epsilon_c$=500 meV and $\omega$=70(100) eV. In all plots, the scale bar is $0.1 \mathrm{\AA}$. }
    \label{ARPES}
\end{figure*}

We suggest that one can use the above spectra to identify each polytype by its unique signatures. We can identify Bernal graphene by the appearance of its third band as stated in Section II. We can see this in the constant momentum cut and -250 meV constant energy cut. To identify the Rhombohedral poly type we see that only one band is present at 0 meV in the constant momentum cut.  Finally, we can distinguish the mixed stacking orders by the intensity of their 3rd and 4th valence bands in the constant energy cuts, in the former the fourth band is clearly visible, particularly at a photon energy of 100 eV. In ABCB the 3rd valence band has a far higher intensity than ABAC, particularly at 100eV photon energy. Finally, we can distinguish the mixed stacking orders by the bright regions in their constant momentum cut. Due to the attenuation of the wave functions from the lower layers, we see the Bernal features more clearly in ABAC's constant momentum cut and similarly the rhombohedral features dominate in ABCB. 

Finally, as in the experimentally fabricated material the number of layer may vary across the flake, we also show (for comparison) the calculated ARPES maps for trilayer graphenes (ABA and ABC) in Fig 2b. We do not compare the 4LG spectra with the spectra of thicker crystals (pentalayers), as their appearance in ARPES is strongly affected by attenuation of electron waves emitted by deeper layers.

\section{Optical conductivity}
Optical conductivity characterises response of a 2D film to light \cite{Kin_evolution_2010, Wang_Polarization_2022,Yu_Gate_2019,kim_chiral_2016}. In particular, its real part describes absorption processes, where energy $\omega$ of an incoming photon is spent in the formation of an electron-hole pair \cite{Kuzmenko_universal_2008}, while the imaginary part gives us insights into the phase shift of the scattered light \cite{ghamsari_measuring_2016}. Because these processes are very sensitive to the band structure as well as the energy and polarisation of the incoming light, optical absorption is widely used to investigate different aspects of graphene systems \cite{mak_optical_2012}, such as identifying the number of layers in graphene stacks \cite{sasaki_universal_2020} or determining the SWMcC parameters \cite{Kuzmenko_determination_2009}. 

In this section, we compute the optical conductivity of graphene stacks with $N$ layers due to inter-band excitations. Denoting $\boldsymbol{l}=(l_x,l_y)$ and $l_z$ the components of the polarisation of light parallel and perpendicular to the sample, the optical conductivity takes the form \cite{Stauber2008,Stauber2013,Ando2009,Tabert2012}
\begin{widetext}
\begin{subequations}
\begin{align}
\mathcal{O}_{\parallel}(\omega)
=&
\frac{i4}{\omega}
\int
\frac{d\boldsymbol{p}}{(2\pi\hbar)^2}
\sum_{n_\mathrm{i}, n_\mathrm{f}}
\frac{
f(\epsilon_{\boldsymbol{p}}^{n_{\mathrm{i}}^-})-
f(\epsilon_{\boldsymbol{p}}^{n_{\mathrm{f}}^+})
}{\omega -(\epsilon_{\boldsymbol{p}}^{n_{\mathrm{f}}^+}-
\epsilon_{\boldsymbol{p}}^{n_{\mathrm{i}}^-})+i\eta^+}
\left|
\bra{\psi_{\boldsymbol{p}}^{n_\mathrm{f}^+}}
e
\boldsymbol{l}
\cdot
\frac{\partial \mathcal{H}}{\partial \boldsymbol{p}}
\ket{\psi_{\boldsymbol{p}}^{n_\mathrm{i}^-}}
\right|^2,\\
\mathcal{O}_{\perp}(\omega)
=&\,
i4\omega
\int
\frac{d\boldsymbol{p}}{(2\pi\hbar)^2}
\sum_{n_\mathrm{i}, n_\mathrm{f}}
\frac{
f(\epsilon_{\boldsymbol{p}}^{n_{\mathrm{i}}^-})-
f(\epsilon_{\boldsymbol{p}}^{n_{\mathrm{f}}^+})
}{\omega -(\epsilon_{\boldsymbol{p}}^{n_{\mathrm{f}}^+}-
\epsilon_{\boldsymbol{p}}^{n_{\mathrm{i}}^-})+i\eta^+}
\left|
\bra{\psi_{\boldsymbol{p}}^{n_\mathrm{f}^+}}
\sigma_0\otimes
\tau
\ket{\psi_{\boldsymbol{p}}^{n_\mathrm{i}^-}}
\right|^2,
\end{align}
\end{subequations}
\end{widetext}

respectively. Above, $\eta^+$ is a positive infinitesimal, $\sigma_0$ the unit matrix in the sublattice space, $\tau\equiv\mathrm{diag}(1,2,\dots,N)ed$, acting on the layer space, is the dipole moment operator and $n_\mathrm{i}^-$ ($n_\mathrm{f}^+$) is the band index of the initial (final) states involved in the electronic transition, respectively.

\begin{figure*}
\centering
\includegraphics[width=1\textwidth]{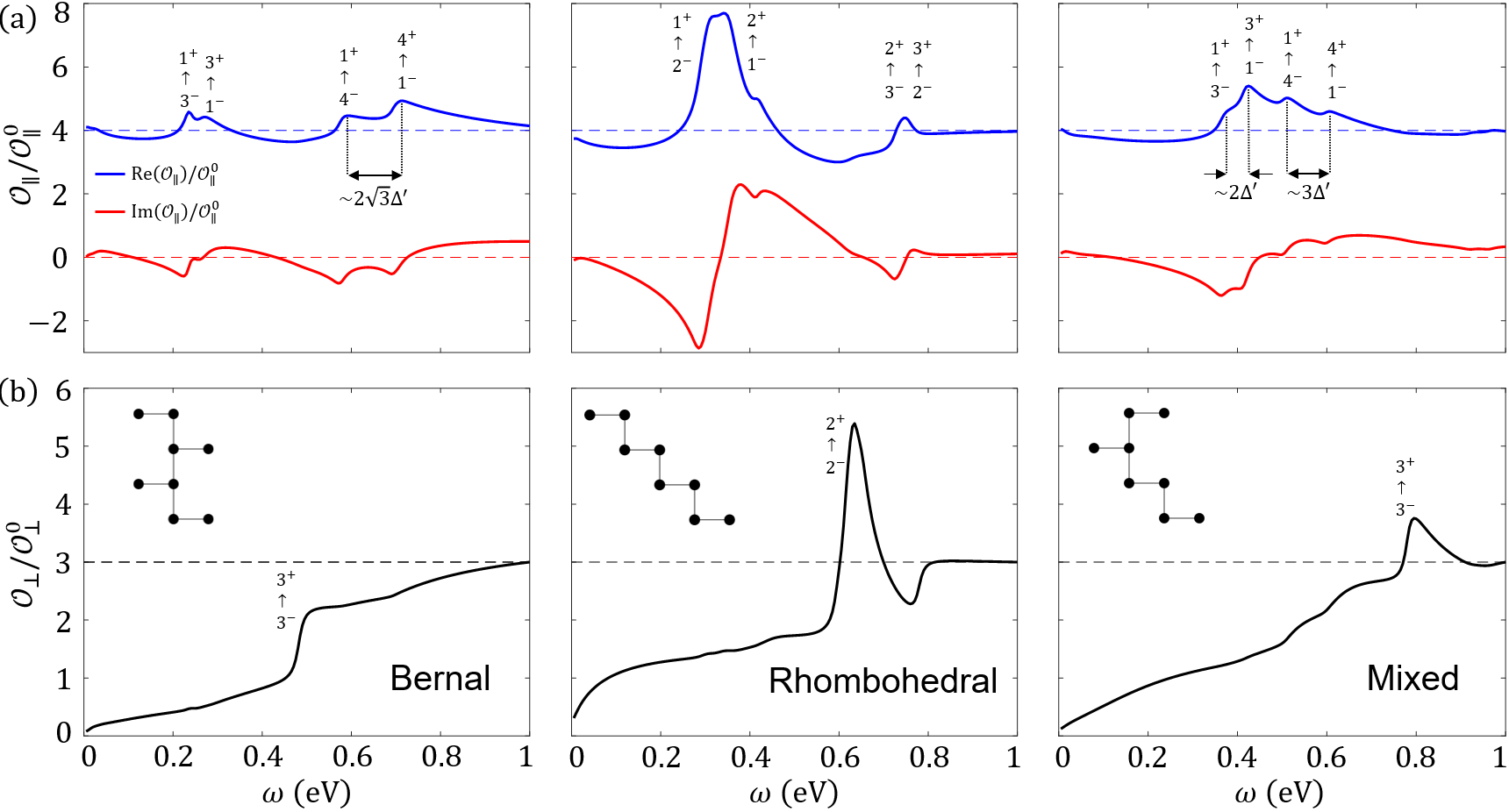}
\caption{ (a) Real and imaginary parts of the in-plane optical conductivity of tetralayer graphenes, in units of the universal optical absorption of graphene, $\mathcal{O}_\parallel^0=\frac{1}{4\pi\varepsilon_0} \frac{\pi e^2}{\hbar c}$. Transitions $1^-\to 3^+,4^+$ are systematically larger than $3^-,4^-\to 1^+$, due to the electron-hole asymmetry and the shift produced by $\Delta'$ in the band structure. (b) Real part of the out-of-plane optical response of tetralayer graphenes, in units of the high-energy out-of-plane response of bilayer graphene, $\mathcal{O}_\perp^0=\frac{c_0^2\gamma_1^2}{2\hbar^2v^2}\,\mathcal{O}_\parallel^0$ \cite{Ando2009}, with an asymptotic value of $3\mathcal{O}_\perp^0$ in all cases. The dominant contribution comes from transitions $n^-\to n^+$.
\label{Fig:Opt_Abs_4LG}}
\end{figure*}

In Fig. \ref{Fig:Opt_Abs_4LG} (a), we present the real and imaginary parts of the in-plane optical conductivity of the three different stacking configurations of tetralayer graphene in units of the optical absorption of graphene, $\mathcal{O}_\parallel^0=
\frac{1}{4\pi\varepsilon_0} \frac{\pi e^2}{\hbar c}$. In all configurations, transitions $n^-\to n^+$ are strongly suppressed, and for photon energies $\omega\sim1$ eV, the absorption spectra are identical and equal to the nominal value of four decoupled graphene layers. In turn, for $\omega<2\gamma_1$, the three absorption spectra exhibit strong differences, which stem from the distinctive features of their corresponding band structure (see Fig. \ref{Fig:band_structures}). In particular, Bernal and ABCB tetralayer graphenes feature four peaks, produced by transitions between the bands close to the Fermi level and the high-energy bands. As mentioned Sec. \ref{Sec:band_structure}, the edges of the split bands, in both the valence and conduction side, are linearly shifted by the parameter $\Delta'$. This results in transitions $1^-\to3^+,4^+$ being systematically more energetic than $3^-,4^-\to1^+$, and provides an experimental route for the determination of $\Delta'$. Conversely, the spectrum of its rhombohedral counterpart is characterised by a sharp peak produced by transitions between the flat bands and the van-Hove singularity in the Mexican-hat bands. 

 In Fig. \ref{Fig:Opt_Abs_4LG} (b), we present the real part of the optical response to perpendicularly polarised light. As opposed to in-plane polarised light, in this case, the main contribution comes from transitions $n^-\to n^+$, thus providing complementary information about energy distances between edges of mirrored bands in the valence and conduction side. For photon energies $\sim1$ eV, all spectra resemble that of three independent graphene bilayers, $3\mathcal{O}_\perp^0$ \cite{Ando2009}, whereas it is again at  mid-infrared frequencies where the absorption spectra give us information about distinct features of the dispersion. In particular, the step-like feature at $0.5$ eV of the Bernal absorption spectrum corresponds to the energy difference between states at the edges of the third bands, and the peaks at $\sim0.6$ and $0.8$ eV in the rhombohedral and ABCB counterparts are linked to transitions that connect the Mexican-hat-shaped bands and heavy-mass bands in the valence and conduction side, as shown in Fig. \ref{Fig:band_structures} (b) and (c), respectively.

\begin{figure*}
\centering
\includegraphics[width=0.9\textwidth]{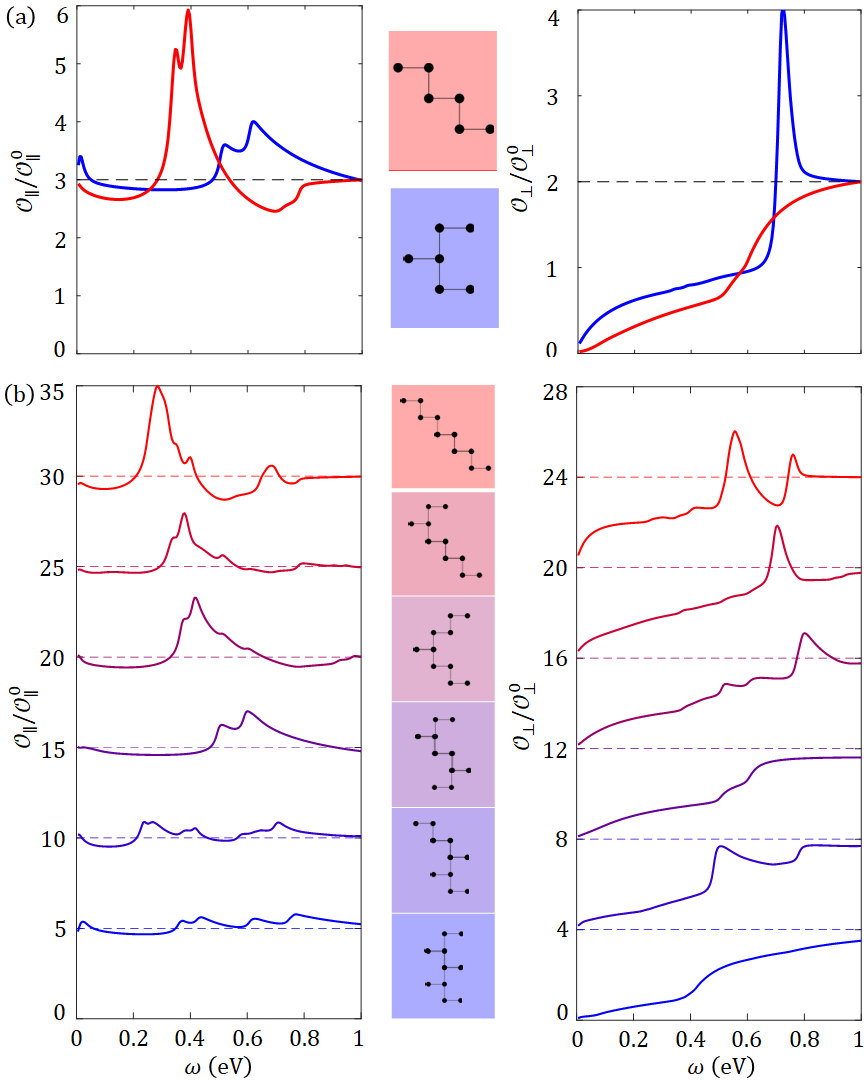}
\caption{ Real part of the in-plane (left) and out-of-plane (right) optical conductivity for all possible configurations of (a) trilayer graphene and (b) pentalayer graphene. For clarity, plots for the parallel and perpendicular optical conductivity of pentalayer graphene are shifted by $5\mathcal{O}_\parallel^0$ and $4\mathcal{O}_\perp^0$, respectively.
\label{Fig:Opt_Abs_3LG_and_5LG}}
\end{figure*}

Finally, owing to the experimental challenge of fabricating films with a given thickness \cite{mak_optical_2012}, we also present the results for 3- and 5-layer films in Figs. \ref{Fig:Opt_Abs_3LG_and_5LG} (a) and (b), respectively, for all possible stacking arrangements. For three-layer films, the in-plane absorption spectra of both rhombohedral and Bernal feature a peak at $0.4$ and $0.6$ eV, split by $\sim2\Delta'$ and $\sim3\Delta'$, respectively. In turn, the most distinctive feature for the out-of-plane absorption spectra is found in the ABC film, a strong peak at $2\gamma_1$. For five layers, as shown in the middle panels of Fig. \ref{Fig:Opt_Abs_3LG_and_5LG} (b), there are six possible stacking orders, as both rhombohedral (top) or Bernal (bottom) may appear with a stacking fault in the outermost layer (second and fifth panels) or in the middle layer (middle two panels). We notice that films with the largest stacking sequences of rhombohedral order feature sharper peaks, and they also lie at lower energies, both in the in-plane and out-of-plane absorption spectrum. This is because the flatness of bands closer to the Fermi level increases with the number of rhombohedral-stacked layers, and the bottoms of the sombrero-like bands also appear at lower energies. This contrasts with the spectra of films with larger stacks in Bernal configuration, which tend to give a featureless constant absorption amplitude.

\section{Electronic Raman scattering}

Raman spectroscopy, based on inelastic scattering of photons, also allows us to get the excitation fingerprint of two-dimensional materials \cite{Liang_Two-dimensional_2018}. In graphene materials, the strongest Raman signals originate from lattice excitations \cite{ferrari_raman_2013}, i.e., phonons, and they provide a wealth of information about the nature of defects \cite{eckmann_probing_2012}, the number of layers \cite{gupta_raman_2006} or presence of strain in the sample \cite{Mohiuddin_Uniaxial_2009}. However, during the last decade, experiments have shown evidence for another type of excitation: the creation of electron-hole pairs \cite{Kashuba_Signature_2009}. While this type of electronic Raman scattering (ERS) produces weaker signals, they come exclusively from the excitation spectra of electrons, and therefore they provide direct information about the band structure of graphene systems, such as the position of van Hove singularities \cite{Aitor_Spectroscopic_2019, Aitor_Electronic_2020} or the formation of gaps \cite{Aitor_Superconductivity_2018}.

\begin{figure*}
    \centering
    \includegraphics[width=1.5\columnwidth]{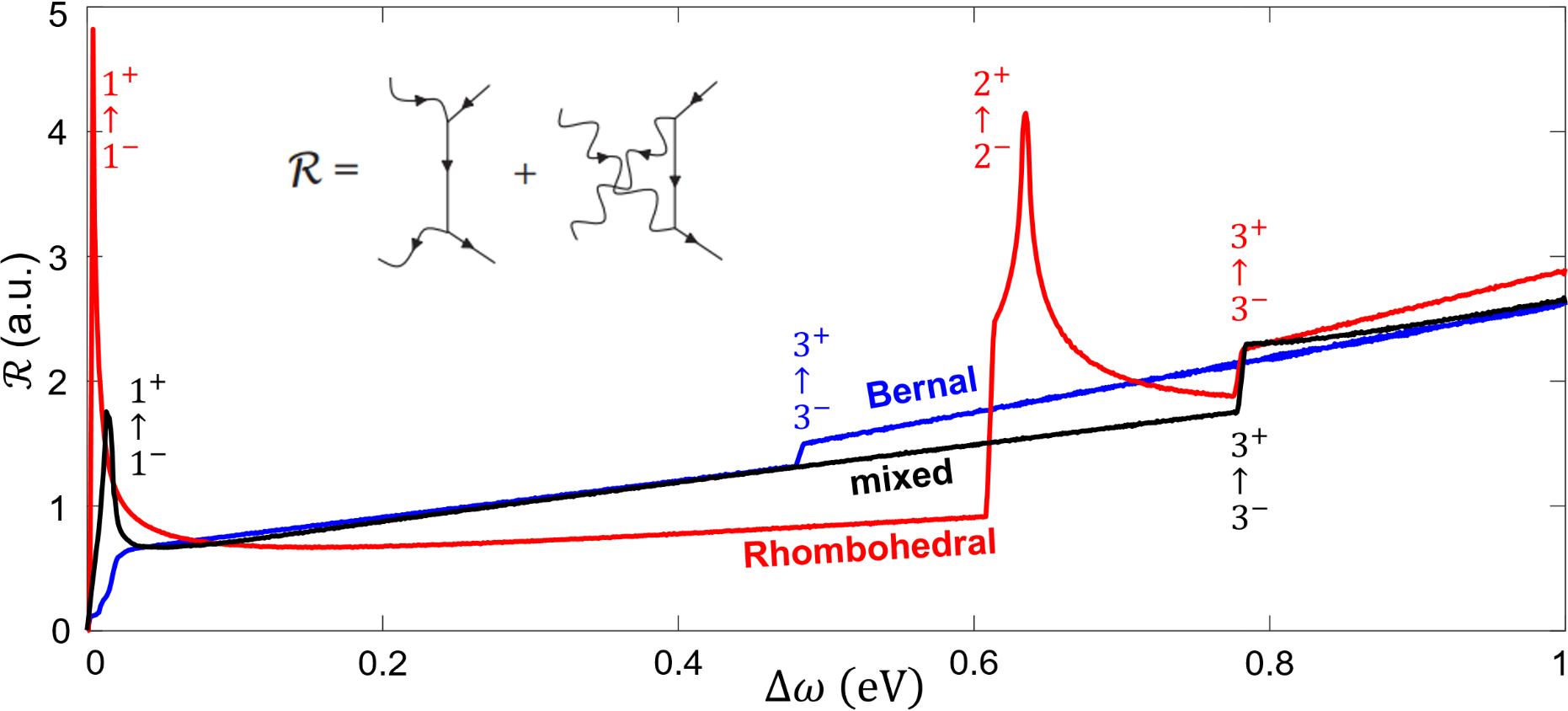}
    \caption{Electronic contribution to Raman scattering in Bernal (blue), rhombohedral (red) and mixed-stacking (black) tetralayer graphene. The inset shows the diagrammatic representation of two-step processes.}
    \label{Fig:ERS}
\end{figure*}

In this section, we model the ERS signals of the three polytypes of tetralayer graphene films presented in Sec. II. In particular, we study the ERS amplitude coming from two-step processes, which were shown to be the dominant contribution in graphene systems \cite{Kashuba_Signature_2009}. This amplitude is given by the evaluation of the two Feynman diagrams in the inset of Fig. \ref{Fig:ERS}, which involve: i) the absorption (emission) at time $t$ of a photon with energy $\omega$ ($\tilde{\omega}$) by an electron from an occupied state in the band with energy $\epsilon_{\boldsymbol{p}}^{n_{\mathrm{i}}^-}$, which is excited into a virtual state and ii) emission (absorption) at time $t'$ of a photon with energy $\tilde{\omega}$ ($\omega$) and one electronic transition into the final state with energy $\epsilon_{\boldsymbol{p}}^{n_{\mathrm{f}}^+}$. The amplitude of such process that results in an electron-hole pair with energy $\Delta\omega\equiv\omega-\tilde{\omega}=\epsilon_{\boldsymbol{p}}^{n_{\mathrm{f}}^+}-\epsilon_{\boldsymbol{p}}^{n_{\mathrm{i}}^-}$ is proportional to
\begin{align}
\mathcal{R}(\Delta\omega)
\propto&
\left|
(\boldsymbol{l}
\times
\tilde{\boldsymbol{l}}^*)_z
\right|^2
\sum_{n_\mathrm{f},n_\mathrm{i}}
\int \frac{d\boldsymbol{p}}{(2\pi\hbar)^2}
[f(\epsilon_{\boldsymbol{p}}^{n_{\mathrm{i}}^-})-
f(\epsilon_{\boldsymbol{p}}^{n_{\mathrm{f}}^+})]
\\
&\times
\left|
\bra{\psi_{\boldsymbol{p}}^{n_\mathrm{f}^+}}
\mathbb{1}_4\otimes
\sigma_z
\ket{\psi_{\boldsymbol{p}}^{n_\mathrm{i}^-}}
\right|^2
\delta(\epsilon_{\boldsymbol{p}}^{n_\mathrm{f}^+}-\epsilon_{\boldsymbol{p}}^{n_\mathrm{i}^-}-\Delta\omega),\nonumber
\end{align}
where $\mathbb{1}_4$ is the $4\times4$ identity matrix acting on the layer space and $\sigma_z$ is the third Pauli matrix acting on the sublattice space. To mention, the prefactor in the equation above determines the selection rules for the polarisation of the scattered light, which needs to be perpendicular to that of the incoming light. This is the hallmark of ERS in graphene materials \cite{Kashuba_Signature_2009,Aitor_Signatures_2020}.

\begin{figure}[h]
    \centering
    \includegraphics[width=0.75\columnwidth]{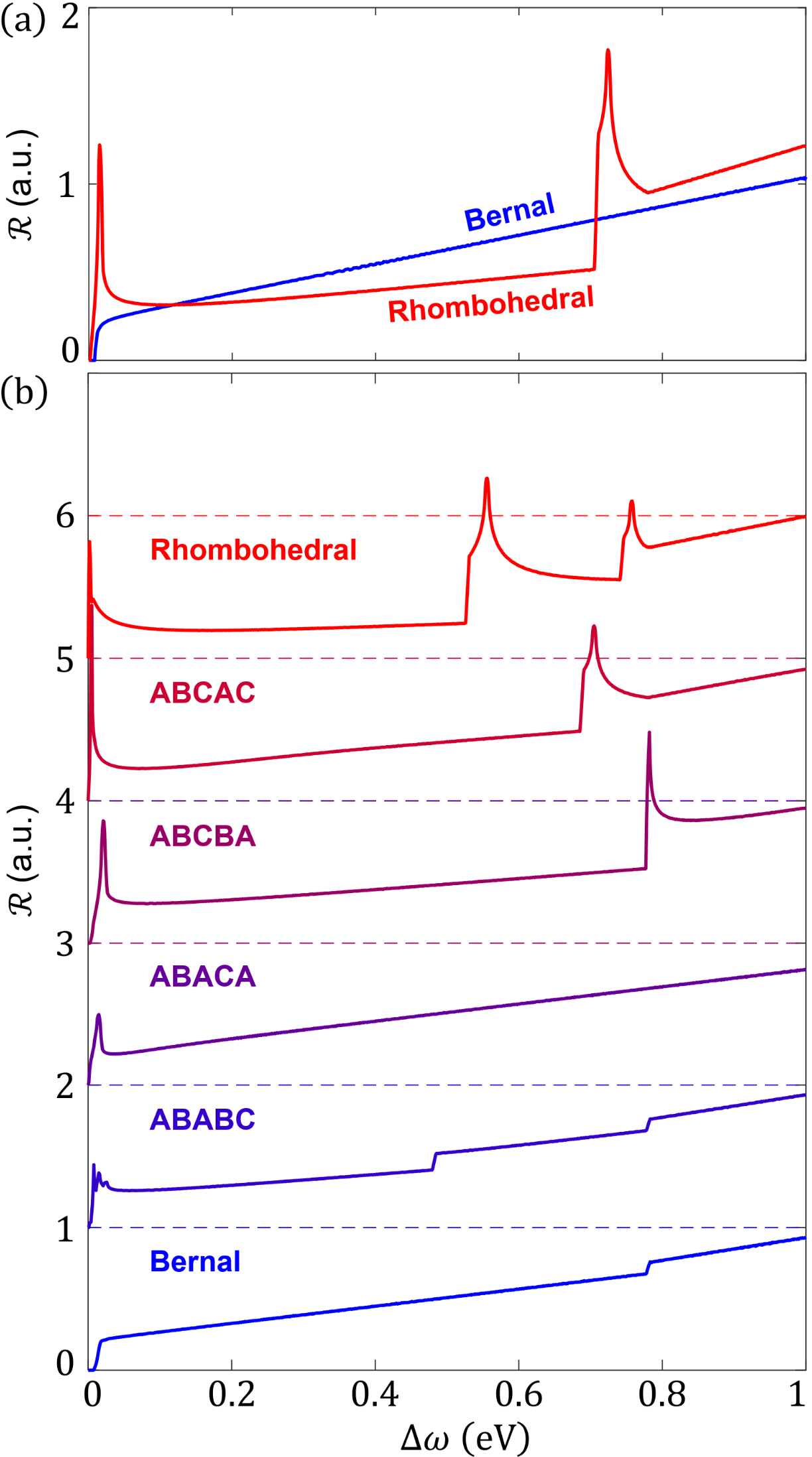}
    \caption{(a) ERS signal for Bernal (blue) and rhombohedral 3LG. (b) ERS signal for the 6 distinct polytypes of 5LG. Each plot is shifted by 1 unit for clarity.}
    \label{Fig:ERS_3LG_5LG}
\end{figure}

The ERS spectra for the three configurations of tetralayer graphene are shown in Fig. \ref{Fig:ERS}. It originated from electronic transitions that connect the $n$ valence band with the $n$ conduction band \cite{Aitor_Signatures_2020}, as opposed to optical absorption where these transitions are forbidden, a common feature of graphene-based systems \cite{Moon2013}. Accordingly, in Bernal tetralayer graphene, we observe a linear trend, with a small step at $2\gamma_1/\varphi \approx500\,\mathrm{meV}$, which corresponds to the energy gap between the parabolic edges of the second conduction and valence bands in Fig. \ref{Fig:band_structures} (a). In contrast, the ERS spectrum of rhombohedral tetralayer graphene features a sharp peak near the origin, due to transitions between flat bands, and at $\sim650\, \mathrm{meV}$, which is the distance between the edges of the sombrero-like bands in Fig. \ref{Fig:band_structures} (b). Finally, we see in the black curve of Fig. \ref{Fig:ERS} that the low energy electronic excitation spectrum of tetralayer films in mixed configuration generate a peak at $\sim15$ meV, followed by a linear trend and a step at $\sim2\gamma_1$, from which the spectra of the three tetralayer films become indistinguishable. Lastly, for comparison, we also present the ERS spectra of graphene films composed of 3 or 5 layers in Fig. \ref{Fig:ERS_3LG_5LG}. For the latter, we observe that films with larger stacks in rhombohedral sequences feature peaks generated by electronic transitions from and to the sombrero-like bands, while parabolic band edges, a common characteristic of films featuring Bernal-stacked layers, result in steps in the ERS intensity. 

\section{Conclusion}
Here we have studied the spectroscopic signatures of the three stacking polytypes of tetralayer graphene. We identify the unique features of each polytype for the different spectroscopic methods investigated. These calculations offer a reference library for the identification of tetralayer graphene polytypes, which offer rich physics - partly due to wide flat intervals in their low-energy bands and partly due to the broken symmetry crystalline structures of mixed-stacking tetralayers. We also note that the exfoliation of tetralayer graphene can produce films where a four-layer system may be adjacent with a thinner or thicker crystal (within the same flake). Therefore, for reference, we also analysed the different stacking polytypes of tri- and pentalayer graphene systems, for these structures to be distinguished from tetralayer graphene. 

\bibliography{Bib}

\section{Appendix A}
In this section to complement the Trilayer and Pentalayer graphene spectra we offer the reader the band structures and density of states plots of each of these systems. All figures shown below were calculated using analogue hybrid $\boldsymbol{k} \cdot \boldsymbol{p}$ theory - tight binding Hamiltonians to those shown in section II above.

\begin{figure*}
    \centering
    \includegraphics[width=\textwidth]{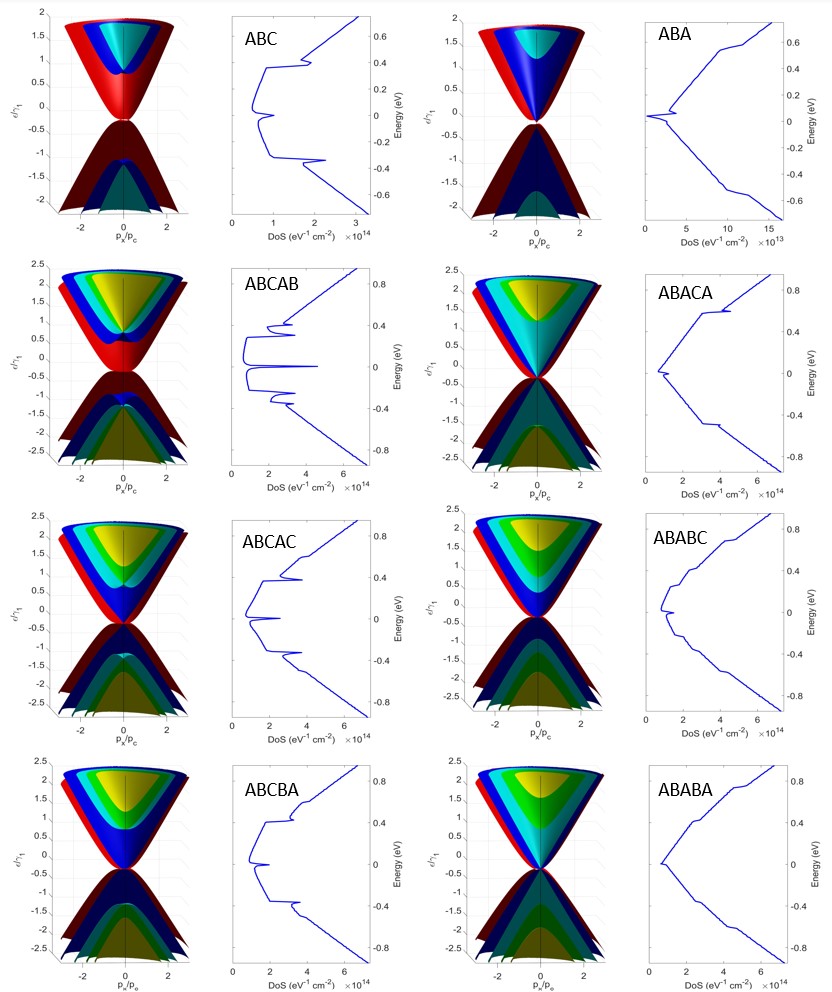}
    \caption{Band structure and density of states of all trilayer and pentalayer graphene polytypes.  In the labels of each panel ${p_c}=\gamma_1/v$ and $\varphi\equiv\frac{1+\sqrt{5}}{2}$ is the golden number.}
    \label{Fig:band_structures_3_5}
\end{figure*}

\end{document}